\documentstyle[prl,aps,amsmath,graphicx]{revtex}
\newcommand{\ave}[1]{\langle #1\rangle}

\begin{document}
%\draft
\twocolumn[\hsize\textwidth\columnwidth\hsize\csname @twocolumnfalse\endcsname
\title{Experimental observation of spatial antibunching of photons}
\author{W. A. T. Nogueira$^{1}$, S. P. Walborn$^{1}$, S.
P\'adua$^{1,2}$, and C. H. Monken$^{1}$\cite{ca}}
\address{$^{1}$Departamento de F\'{\i}sica, Universidade Federal de
Minas Gerais, Caixa Postal 702, Belo Horizonte, MG 30123-970, Brazil\\
$^{2}$Dipartimento di Fisica, Universit\`a ``La Sapienza'', Roma, 00185, Italy}
\date{\today}
\maketitle
\begin{abstract}
We report an interference experiment that shows transverse spatial
antibunching of photons. Using collinear parametric down-conversion
in a Young-type fourth-order interference setup we show interference
patterns that violate classical Schwarz inequality and should not
exist at all in a classical description.
\end{abstract}
]
\thispagestyle{empty}
Photon antibunching in a stationary field is recognized as a signature
of nonclassical behavior, for its description is not possible in terms
of a nonsingular positive Glauber-Sudarshan $P$ distribution
\cite{mandel95}. It is well known that any state of the
electromagnetic field that has a classical analog can be described by
means of a positive $P$ distribution which has the properties of a
classical probability functional over an ensemble of coherent states.

The classical intensity correlation function for stationary fields must
obey the following inequality \cite{mandel95}:
\begin{equation}
\ave{I({\mathbf r},t)I({\mathbf r},t+\tau)} \leq \ave{I^{2}({\mathbf r},t)}.
\label{in1}
\end{equation}
All field states described in terms of a positive nonsingular $P$
distribution must obey the standard quantum mechanical counterpart of
(\ref{in1}), where products of intensities are replaced by ordered
products of photon density operators \cite{mandel95}, that is,
\begin{equation}
\ave{{\mathcal T\!}:\hat{I}({\mathbf r},t)\hat{I}({\mathbf r},t+\tau):} \leq
\ave{:\hat{I}^{2}({\mathbf r},t):},
\label{eq:in1q}
\end{equation}
where ${\mathcal T\!}:\  :$ stands for time and normal ordering.
Photon density operators are defined as
\begin{equation}
\hat{I}({\mathbf r},t)=\hat{\mathbf V}^{\dagger}({\mathbf
r},t)\hat{\mathbf V}({\mathbf r},t),
\end{equation}
where
\begin{displaymath}
\hat{\mathbf V}({\mathbf r},t)=\sum_{{\mathbf k},\sigma}
\hat{a}_{{\mathbf k},\sigma}\bbox{\epsilon}_{{\mathbf k},\sigma}
e^{i({\mathbf k}\cdot{\mathbf r}-\omega t)},
\end{displaymath}
$\hat{a}_{{\mathbf k},\sigma}$ is the annihilation operator for the
mode with wave vector ${\mathbf k}$ and polarization $\sigma$,
$\bbox{\epsilon}_{{\mathbf k},\sigma}$ is the unit polarization
vector, and $\omega = c k$.

Inequality (\ref{eq:in1q}) means that for such class of fields,
photons are detected either bunched or randomly distributed in time.
Photon antibunching in time, characterized by the violation of
(\ref{eq:in1q}), was predicted by Carmichael and Walls
\cite{carmichael76}, Kimble and Mandel \cite{kimble76}, and was first
observed by Kimble, Dagenais and Mandel in resonance fluorescence
\cite{kimble77}.

Let us now turn to space domain and consider that the transverse field
profile of a given stationary light beam propagating along $z$
direction is described by a complex stochastic vector amplitude
$\bbox{{\mathcal V}}(\bbox{\rho},t)$ with an associated probability
functional ${\mathcal P}(\bbox{{\mathcal V}})$.  Here, $\bbox{\rho}$
lies in a plane transverse to the propagation direction.  The average
intensity at a point $\bbox{\rho}$ is
\begin{equation}
\ave{ I(\bbox{\rho},t )} = \ave{\bbox{\mathcal
V}^{*}(\bbox{\rho},t)\bbox{\mathcal V}(\bbox{\rho},t)} =
\int {\mathcal P}({\bbox{{\mathcal
V}}}) |\bbox{\mathcal V}(\bbox{\rho},t)|^{2}\
d{\bbox{{\mathcal V}}},
\label{av1}
\end{equation}
and the two-point intensity correlation function
\begin{displaymath}
\Gamma^{(2,2)}(\bbox{\rho}_{1},\bbox{\rho}_{2},\tau)=
\ave{I(\bbox{\rho}_{1},t)I(\bbox{\rho}_{2},t+\tau)}
\end{displaymath}
is
\begin{equation}
\Gamma^{(2,2)}(\bbox{\rho}_{1},\bbox{\rho}_{2},\tau)=\int
{\mathcal P}(\bbox{\mathcal V}) |\bbox{\mathcal
V}(\bbox{\rho}_{1},t_{1})|^{2}|\bbox{\mathcal
V}(\bbox{\rho}_{2},t_{2})|^{2}\ d\bbox{\mathcal V}.
\label{av2}
\end{equation}
Its time dependence is restricted to the difference $\tau=t_{1}-t_{2}$,
since the field is assumed to be stationary.  In the space domain, the
concept analogous to stationarity is homogeneity.  For a homogeneous
field, the expectation value of any quantity that is a
function of position  is invariant under translation of the
origin \cite{mandel95}. In particular,
\begin{equation}
\Gamma^{(2,2)}(\bbox{\rho}_{1},\bbox{\rho}_{2},\tau)=
\Gamma^{(2,2)}(\bbox{\delta},\tau)
\label{eq:stat1}
\end{equation}
and
\begin{equation}
\ave{I^{N}(\bbox{\rho}+\bbox{\delta},t+\tau)} =
\ave{I^{N}(\bbox{\rho},t)},
\label{eq:stat2}
\end{equation}
where $\bbox{\delta}=\bbox{\rho}_{1}-\bbox{\rho}_{2}$ and $N=1,2,\ldots$

Applying Schwarz inequality,
\begin{equation}
\ave{I(\bbox{\rho},t)I(\bbox{\rho}+\bbox{\delta},t+\tau)}^{2}
\leq \ave{I^{2}(\bbox{\rho},t)}
\ave{I^{2}(\bbox{\rho}+\bbox{\delta},t+\tau)}.
\label{eq:cs1}
\end{equation}
By means of (\ref{eq:stat2}),
\begin{equation}
\ave{I(\bbox{\rho},t)I(\bbox{\rho}+\bbox{\delta},t+\tau)}
\leq \ave{I^{2}(\bbox{\rho},t)}.
\label{eq:cs2}
\end{equation}
Quantum mechanically,
\begin{equation}
\ave{{\mathcal 
T\!}:\hat{I}(\bbox{\rho},t)\hat{I}(\bbox{\rho}+\bbox{\delta},t+\tau):}
\leq \ave{:I^{2}(\bbox{\rho},t):},
\label{eq:cs2q}
\end{equation}
that is,
\begin{equation}
\Gamma^{(2,2)}(\bbox{\delta},\tau)\leq
\Gamma^{(2,2)}(\bbox{0},0).
\label{eq:bunch}
\end{equation}
Analogously to what was concluded from inequality (\ref{eq:in1q}), for
field states represented by positive nonsingular Glauber-Sudarshan
distributions, that is, fields that admit the classical stochastic
description assumed above, inequality (\ref{eq:cs2q}) implies that
photons are detected either spatially bunched or randomly spaced in a
transverse detection screen.  Spatial antibunching of photons has been
predicted by some authors
\cite{rousseau79,klyshko82,birula91,kolobov91,kolobov99} and a
possible experiment was recently proposed to observe it in squeezed
states \cite{kolobov91,kolobov99}.

In this work we show that strong antibunching in one transverse
direction can be observed in down-converted light, violating
(\ref{eq:cs2q}) by several standard deviations.  The effect is
produced by fourth-order interference of a two-photon beam diffracted
by a birefringent double-slit.  The experimental setup is depicted in
Fig.  1.  A light beam of $\lambda = 702$\,nm is produced by collinear
type II down-conversion in a 2\,mm-long nonlinear crystal ($BBO$)
pumped by an Argon laser beam with $\lambda = 351$\,nm.  The u.\,v.
beam transmitted by the crystal is removed from the down-converted
beam by a laser mirror ($M$) transparent to 702\,nm.  A birefringent
double slit ($S$) is constructed as follows.  A single slit of
dimensions 0.60\,mm$\times$5\,mm is divided in two by a 0.20\,mm wide
absorbing strip, defining two parallel slits of dimensions
0.20\,mm$\times$5\,mm.  In front of each slit there is a quarter wave
plate ($Q_{1}$ and $Q_{2}$), as shown in Fig.  2.  One wave plate
($Q_{1}$) has its fast axis parallel to the slits, whereas the other
one ($Q_{2}$), has its fast axis perpendicular to the slits.  With
such alignment, the waveplates introduce a phase difference of $\pi$
between the two slits.  This arrangement is placed in the
down-converted beam, 38\,cm from the crystal.  The pumping beam is
focused right on the plane of the double slit by a lens ($L$) of
500\,mm focal length.  Assuming that the beams are propagating along
the $z$ direction, this focusing causes the fourth-order correlation
function of the down-converted beam to be concentrated on points
satisfying $\bbox{\xi}_{1}+\bbox{\xi}_{2}=0$, where
$\bbox{\xi}_{1}$ and $\bbox{\xi}_{2}$ are position vectors on the
plane of the double slit \cite{monken98}.  The focusing is essential
to produce the appropriate spatial dependence of the diffracted field
\cite{monken00}.  In order to make possible that two detectors
($D_{1}$ and $D_{2}$) share the same transverse position without being
limited by their physical dimensions, a beam splitter ($BS$) is
inserted in the down-converted beam, with $D_{1}$ and $D_{2}$ placed
in front of each exit port.  In front of each detector, there is a
single slit of dimensions 0.20\,mm$\times$3\,mm aligned horizontally
(parallel to the slits in $S$), followed by an interference filter
with a bandwidth of 40\,nm, centered at 690\,nm, and a
lens focused on the detector's active area.  The optical path length from
the double slit $S$ to the detectors $D_{1}$ and $D_{2}$ is 70\,cm.
$D_{1}$ and $D_{2}$ are mounted on precision translation stages and
their vertical positions are set by computer-controlled stepping
motors.  Single and coincidence counts were measured while $D_{1}$ and
$D_{2}$ were scanned in the vertical direction ($x$ axis), as will be
described below.

Ideally, the transverse fourth-order correlation function
$\Gamma^{(2,2)}(\bbox{\rho},0)$ is proportional to the coincidence
rate between two punctual detectors separated by $\bbox{\rho}$, with a
negligible resolving time.  Since the detectors are not punctual and
the coincidence resolving time is finite (10\,ns in our setup), what
was actually measured is a convolution of
$\Gamma^{(2,2)}(\bbox{\delta},\tau)$ with the sampling window $\Delta
x\Delta y\Delta\tau$, where $\Delta x$ and $\Delta y$ represent the
dimensions of the detector entrance slit (0.20\,mm$\times$3\,mm) and
$\Delta\tau$ is the resolving time of the coincidence counter
(10\,ns).  For the purpose of demonstrating the effect, however, we
will ignore this correction by considering $\Delta x\simeq 0$, $\Delta
y \rightarrow \infty$, and $\Delta \tau \simeq 0$.  Under these
conditions, it is possible to show \cite{monken00} that for small
displacements, the coincidence rate is proportional to
\begin{equation}
1-\cos\left[\frac{2\pi d}{\lambda z}(x_{2}-x_{1})\right],
\label{eq:interf}
\end{equation}
where $\lambda$ is the wavelength of the down-converted field
(702\,nm), $d$ is the double slit separation (0.40\,mm), $z$ is the
optical path length between the double slit and the detectors
(70\,cm), $x_{1}$ and $x_{2}$ are the vertical positions of detectors
$D_{1}$ and $D_{2}$, respectively.

Before taking correlation measurements, the accuracy
of vertical positioning was checked by the following procedure.  With
the double slit $S$ removed, a horizontally aligned wire was stretched
in front of the beam splitter, at $x=0$, and single counts were
registered in sampling times of 5\,s, while the detectors were scanned
vertically.  The result is shown in Fig.  3.  The two counting
profiles are not identical due to differences in the overall quantum
efficiencies of $D_{1}$ and $D_{2}$.

Figures 4 to 7 summarize the results of single counts and coincidence
measurements taken in sampling times of 1000\,s in several different
situations.  All coincidence patterns were fit to expression
(\ref{eq:interf}) plus a background.  Vertical error bars are
statistical with two standard deviations in length, whereas horizontal
ones correspond to the width of detectors entrance slits.  The results
shown in Fig.  4 refer to the situation in which detector $D_{2}$ is
kept at $x_{2}=0$ and detector $D_{1}$ is scanned vertically.  The
single counts, although not constant, do not show any oscillation to
which one could attribute the oscillation in coincidences.  The same
is true in Fig.  5, where detector $D_{1}$ was kept in $x_{1}=0$ and
$D_{2}$ was scanned vertically.  When $D_{1}$ and $D_{2}$ were scanned
together $(x_{1}=x_{2})$, a fairly constant background of coincidences
was recorded, as shown in Fig.  6. This background, that should be zero as
well as the minima in Fig.  4 and Fig.  5, is due to the finite width
of the detectors entrance slits (0.20\,mm).  A final measurement was
performed by scanning $D_{1}$ with $D_{2}$ kept in the position
$x_{2}=-0.55$\,mm, which corresponds to a maximum of coincidences in
Fig.  5.  The results are plotted in Fig.  7, showing that the minimum
in coincidences was displaced to $x_{1}=x_{2}=-0.55$\,mm.

All the interference patterns shown here satisfy
\begin{equation}
\Gamma^{(2,2)}(\bbox{\delta},0) >
\Gamma^{(2,2)}(\bbox{0},0),
\label{eq:antib}
\end{equation}
in a clear violation of expression (\ref{eq:bunch}), characterizing
the presence of transverse spatial antibunching of photons.

Let us analyze these results from another point of view.  Some years
ago, it was pointed out \cite{klyshko92} that all second- and
fourth-order optical interference effects observed so far have close
classical analogs with the same harmonic pattern, differing only in
their visibilities.  This is not the case of our results.  Since the
minimum of fourth-order interference occurs for $x_{1}=x_{2}$
in the absense of second-order interference, any classically
predicted visibility different from zero would violate Schwarz
inequality.  Therefore, our results can be regarded as a truly quantum
fourth-order interference effect.

The authors acknowledge the support from the Brazilian agencies
CNPq, FINEP, PRONEX, and FAPEMIG. S. P\'adua acknowledges CAPES for a Scholar
fellowship at Universit\`a ``La Sapienza''.

%%%%%%%%%%%%%%%%%%%%%%%%%%%%%%%%%%%%%%%%%%%%%%%%%%%%%%%%%%%%%%%%%%%%%%%%%%%%%%%%%
%                                      Figures 
%
%%%%%%%%%%%%%%%%%%%%%%%%%%%%%%%%%%%%%%%%%%%%%%%%%%%%%%%%%%%%%%%%%%%%%%%%%%%%%%%%%
\begin{figure}
\centerline{\includegraphics{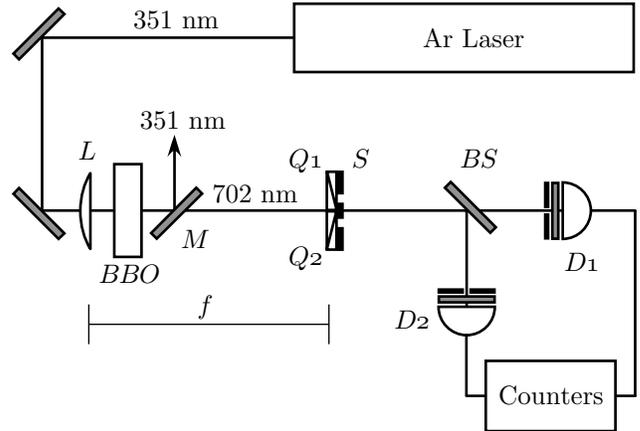}}
\caption{Experimental setup.  $L$ is a lens of focal length
$f=500$\,mm, $BBO$ is a 2\,mm-long $\beta$-BaB$_{2}$O$_{4}$ nonlinear
crystal cut for collinear type II 351\,nm$\rightarrow$702\,nm
down-conversion, $M$ is a u.\,v.  high reflectance mirror, $Q_{1}$ and
$Q_{2}$ are quarter-wave plates, $S$ is a double slit, $BS$ is a 50:50
beam splitter, $D_{1}$ and $D_{2}$ are avalanche photo-diodes working
in photon counting mode.}
\end{figure}
\begin{figure}
\centerline{\includegraphics{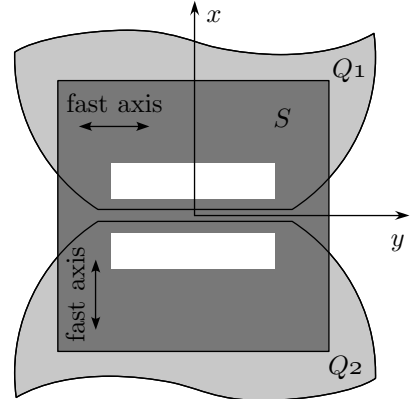}}
\caption{The birefringent double slit. $Q_{1}$ and $Q_{2}$ are
quarter-wave plates aligned with orthogonal fast axes, and $S$ is a
double slit with clear apertures of 0.20\,mm$\times$5\,mm separated by a
0.20\,mm obstacle.}
\end{figure}
\newpage
\begin{figure}
\centerline{\includegraphics[scale=0.9]{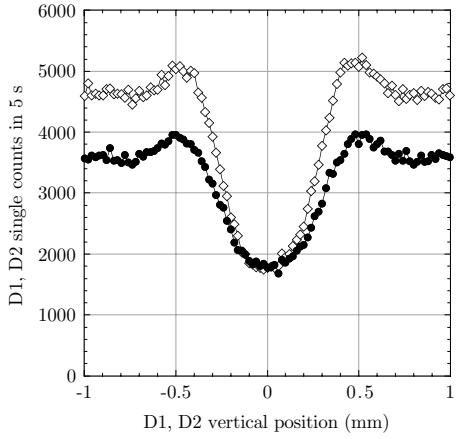}}
\caption{$D_{1} (\diamond)$ and $D_{2} (\bullet)$ single counts taken
with a 0.20\,mm diameter wire stretched horizontally in front of the
beam splitter and the double slit removed. This measurement was taken
in order to check the accuracy of detectors vertical positioning.}
\end{figure}
\begin{figure}
\centerline{\includegraphics[scale=0.9]{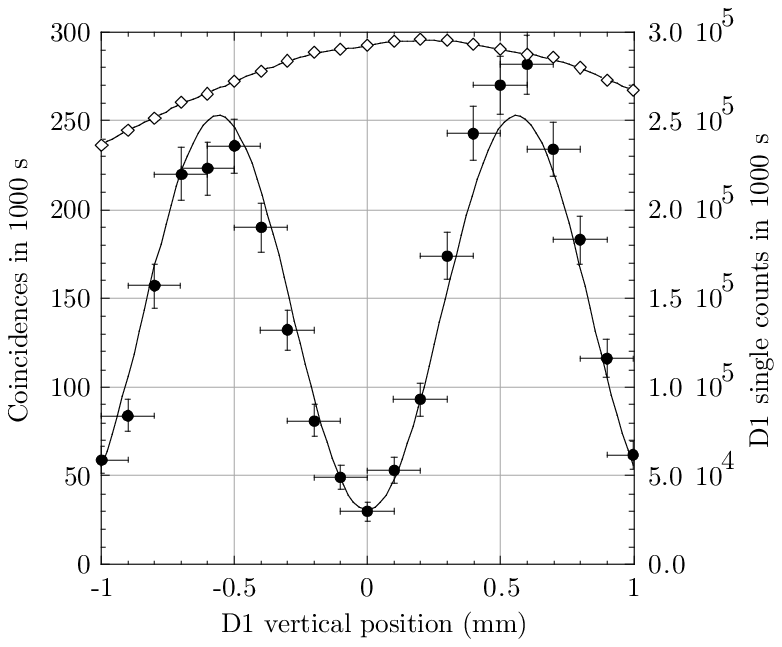}}
\caption{Single counts ($\diamond$) and coincidences ($\bullet$) taken with
$D_{2}$ kept in $x_{2}=0$ and $D_{1}$ scanned vertically.}
\end{figure}
\begin{figure}
\centerline{\includegraphics[scale=0.9]{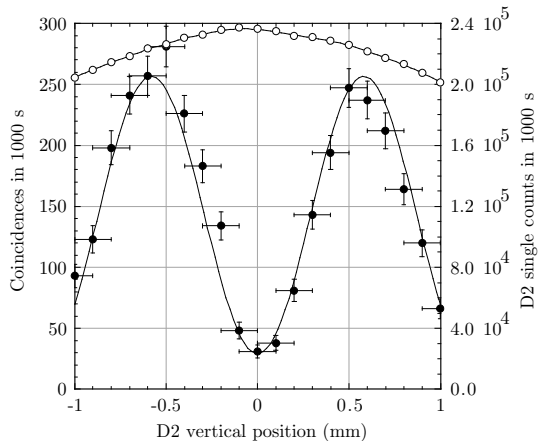}}
\caption{Single counts ($\circ$) and coincidences ($\bullet$) taken with
$D_{1}$ kept in $x_{1}=0$ and $D_{2}$ scanned vertically.}
\end{figure}
\begin{figure}
\centerline{\includegraphics[scale=0.9]{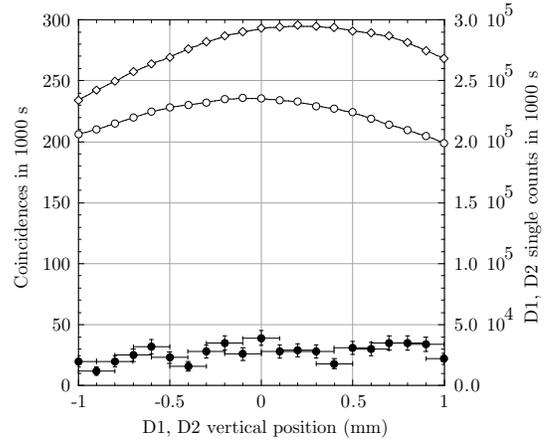}}
\caption{$D_{1}$ single counts ($\diamond$) , $D_{2}$ single counts
$(\circ)$, and coincidences ($\bullet$) taken when both $D_{1}$ and $D_{2}$
were scanned vertically, keeping $x_{1}=x_{2}$.}
\end{figure}
\begin{figure}
\centerline{\includegraphics[scale=0.9]{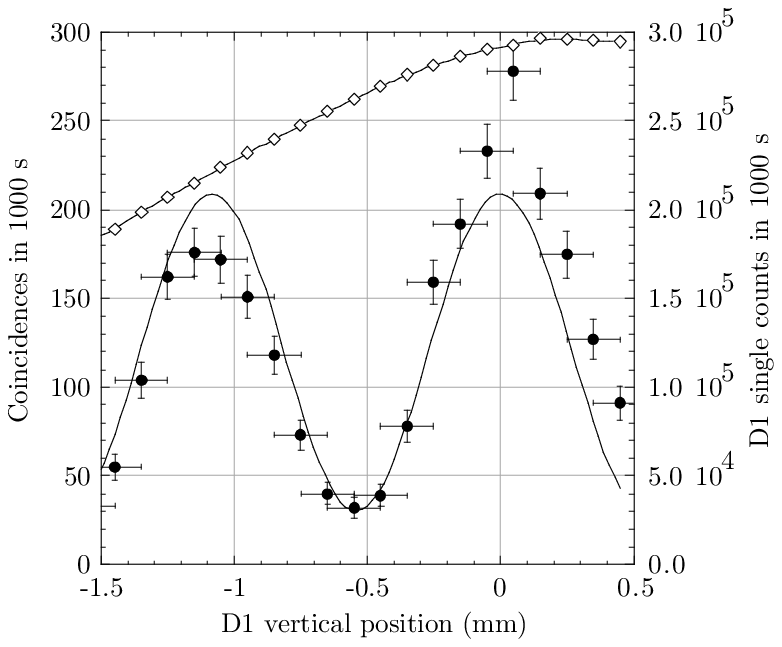}}
\caption{Single counts ($\diamond$) and coincidences ($\bullet$) taken with
$D_{2}$ kept in $x_{2}=-0.55$\,mm and $D_{1}$ scanned vertically.}
\end{figure}
\end{document}